\documentclass[preprint,12pt]{elsarticle}
\usepackage{enumitem}
\usepackage{amssymb}
\usepackage{xcolor}        
\usepackage{floatpag}      
\usepackage{chemformula}   

\begin{document}

\begin{frontmatter}

\title{Single-photon photoionization of oxygen-like Ne III}

\author[1]{S. N. Nahar} 
\author[2]{A. M. Covington}
\author[3]{D. Kilcoyne}
\author[2]{V. T. Davis}
\author[2]{J. F. Thompson}
\author[4]{E. M. Hern\'andez}
\author[4]{A. Antill\'on}
\author[4]{A. M. Ju\'arez}
\author[4]{A. Morales-Mori}
\author[4]{G. Hinojosa\thanks{hinojosa@icf.unam.mx}}

\address[1]{Department of Astronomy, The Ohio State University, OH 43210-1173, USA}
\address[2]{Physics Department, University of Nevada, Reno NV 89557-0220, USA}
\address[3]{The Advanced Light Source, Lawrence Berkeley National Laboratory, CA 94720, USA}
\address[4]{Instituto de Ciencias F\'{i}sicas, Universidad Nacional Aut\'onoma de M\'exico,
       A. P. 48-3, Cuernavaca 62251, Mexico}
       
\journal{International Journal Mass Spectrometry}

\begin{abstract}
 We offer a theoretical and experimental study of the single-photon photoionization 
 of Ne {\footnotesize III}. The high photon flux and the high-resolution capabilities
 of the Advanced Light Source at the LBNL 
 were employed to measure absolute photoionization cross sections.
 The resulting spectrum has been benchmarked against high accuracy relativistic 
 Breit-Pauli $R$-matrix calculations. A large close-coupling wave function
 expansion which comprises up to 58 fine-structure levels of the residual 
 ion Ne {\footnotesize IV} of configurations $2s^22p^3$, $2s2p^4$, $2p^5$, 
 $2s^22p^23s$, $2s^22p^23p$ and $2s^22p^23d$ was included. 
 A complete identification of the measured features was achieved by considering
 seven low-lying levels of Ne {\footnotesize III}. We found that the photoionization 
 cross-section ($\sigma_{PI}$) exhibits the presence of prominent resonances 
 in the low-energy region near the ionization thresholds that correspond to 
 low-lying levels. These include high-peak narrow resonances with almost 
 zero background introduced by relativistic effects. However, there 
 does not exist a significant contribution to  $\sigma_{PI}$ from relativistic effects 
 at the high-energy interval of the present study.
\end{abstract}


\end{frontmatter} 

\section{Introduction}

 There is a growing interest in advancing the present knowledge about the
 photoionization (PI) of ions due to the wide range of applications in which 
 the interaction of ions with radiation is of relevance. For example, in 
 plasma science, autoionizing states induced during PI are proposed as a 
 source of thermal electrons \cite{bartnik2015}. Knowledge of 
 the autoionization energies is critical in the derivation of charge-state 
 distributions in plasmas \cite{ralchenko2016}. This knowledge is of 
 importance in the  opacity problem which is paramount in astrophysics 
 \cite{bailey2015}. The search for life-propitious regions in the Universe 
 \cite{koo2013}, the study of star-formation \cite{jaskot2016} and the 
 understanding of certain atmospheric environments \cite{thissen2011} rely 
 also, at least in part, on the ability to generate atomic models of PI 
 to help unravel the complex problem of identifying atomic species and 
 determining their abundances.
 
 Although, significant progress has been made in understanding PI processes 
 for several elements \cite{kjeldsen2006,schippers2016,west2001}, neon 
 ions have been the subject of relatively few PI studies, despite the fact 
 that the presence of \ch{Ne} has been confirmed in planetary nebulae 
 \cite{shure1984}, and that \ch{Ne} is one of the most abundant elements in 
 the Universe.
 
 In a theoretical approach based on the close-coupling approximation by 
 Pradhan \cite{pradman1979,pradman1980} the single-photon PI cross-sections 
 for the ground states of Ne{\footnotesize \ II}, Ne{\footnotesize \ III} and 
 Ne{\footnotesize\ IV} were calculated. Detailed cross-sections for a large 
 number of states of Ne{\footnotesize \ III} were calculated using the
 non-relativistic $R$-matrix method in the $L \cdot S$ coupling approximation
 by Butler and Zeippen under the auspices of the Opacity  Project (OP) 
 \cite{op} (data sets are available online \cite{topbase2018}).
 
 With the aid of the $R$-matrix method it has been possible, under the 
 $L\cdot S$ coupling approximation, to construe most of the observed features 
 and cross-section intensities in high-resolution, single-photon PI measurements 
 of Ne{\footnotesize \ II} \cite{covington2002} and Ne{\footnotesize \ IV} 
 \cite{aguilar2005} cations.
 
 This work is concerned with the photoionization of $L$-shell electrons
 of \ch{Ne}, however, it is important to point out that studies of the inner 
 shell and near $K$-edge photoionization of Ne{\footnotesize \ I} 
 \cite{PhysRevA.70.062502}, Ne{\footnotesize \ II} \cite{PhysRevA.65.012709} 
 \cite{taj2017} are available. Isoelectronic to \ch{Be}, Ne{\footnotesize \ VII}
 photoionization have also been provided \cite{PhysRevA.66.022715, Kim_2012}.  
 
 Ne{\footnotesize \ III} is isoelectronic to O{\footnotesize \ I}, so 
 previous studies on the photoionization of the oxygen atom are relevant
 to the interpretation of the present data. Angel and Samson \cite{angel1988} 
 (and references therein) identified several autoionizing states leading to 
 O{\footnotesize \ II} and measured the relative single-photon PI cross section 
 with the help of a monochromator using radiation from a synchrotron. 
 In a calculation that included core excited states,  Nahar \cite{nahar1998}
 obtained the PI cross-section of the isonuclear series O{\footnotesize \ I} 
 to O{\footnotesize \ VII} using an {\it ab-initio} $R$-matrix method.
 Flesh and collaborators \cite{flesh2000} investigated the laser-induced, 
 single-photon ionization of excited $^1D_2$ state of O{\footnotesize \ I};
 they found Rydberg series' below the ground state of O{\footnotesize \ II}.
 Tayal \cite{tayal2002} calculated the single-photon PI cross section for the
 ground state O{\footnotesize \ I} using an $R$-matrix method.
 More recently, relevant parameters for states lying over the ionization 
 limit of  Ne{\footnotesize \ III} have been studied \cite{Eser2016}.

 In this paper, we offer  
 high-resolution measurements of the single-photon photoionization  
 of the Ne{\footnotesize\ III} cation and an interpretation 
 of the resulting spectra based on {\it ab-initio} 
 coupled-channel (CC) Breit-Pauli $R$-matrix (BPRM) calculations. Energies for 
 resonant structures and possible assignments are also provided.
 
\section{Theory}

 We have used the close-coupling approximation in which the atomic system 
 is represented by (N+1) electrons. N is the number of core electrons in 
 the residual ion interacting with the (N+1)$^{th}$ electron. The 
 (N+1)$^{th}$ electron is bound or in the continuum depending on whether 
 its energy (E) is negative or positive . The total wave function $\Psi_E$, 
 with symmetry $J\pi$, is expressed by the expansion (for details, for example, 
 see \cite{aas})

 \begin{equation}
   \Psi_E(e^- + ion) = A \sum_{i} \chi_{i}(ion)\theta_{i} + \sum_{j} c_{j} \Phi_{j} \ ,
 \end{equation}

 \noindent
 where the core ion eigenfunction, $\chi_{i}$ of state $S_iL_i~J_i\pi_i$, 
 represents ground and various excited states, and the sum is over the number 
 of considered states. The core is coupled with the wave function $\theta_{i}$ 
 of the (N+1)$^{th}$ electron which has kinetic energy $k_{i}^{2}$ and orbital 
 angular momentum $\ell_i$ in a channel labeled as 
 $S_iL_i J_i\pi_ik_{i}^{2}\ell_i(SL J\pi)$. $A$ is the antisymmetrization operator. 
 In the second sum, the $\Phi_j$s are bound functions of the (N+1)-electron 
 system that provide for the orthogonality between the continuum and the 
 bound electron orbitals, and account for short-range correlation. When core 
 excitations are included in the wave function in any {\it ab initio} 
 calculation, as in the close coupling (CC) approximation, resonances are 
 inherently generated. Substitution of $\Psi_E(e^- + ion)$ into the Schr\"odinger 
 equation
\begin{equation}
H_{N+1}\mit\Psi_E = E\mit\Psi_E
\end{equation}

\noindent
 introduces a set of coupled equations that are solved using the $R$-matrix
 approach. The details of the $R$-matrix method in the CC approximation can 
 be found in e.g., \cite{aas,br75,mjs87,betal87,betal95}. 

 Relativistic effects are included via the Breit-Pauli approximation wherein 
 the Hamiltonian is given by

\begin{eqnarray}
H_{N+1}^{\rm BP}= \sum_{i=1}\sp{N+1}\left\{-\nabla_i\sp 2 - \frac{2Z}{r_i}
 + \sum_{j>i}\sp{N+1} \frac{2}{r_{ij}}\right\} +\nonumber
\end{eqnarray}
\begin{eqnarray}
H_{N+1}^{\rm mass} +
H_{N+1}^{\rm Dar} + H_{N+1}^{\rm so} \,
\end{eqnarray}

\noindent
 in units of Rydberg. The Darwin relativistic terms and mass corrections are  
 $H^{\rm Dar} = {Z\alpha^2 \over 4} \sum_i{\nabla^2({1 \over r_i})}$ and 
 $H^{\rm mass} = -{\alpha^2\over 4}\sum_i{p_i^4}$.  The spin-orbit interaction 
 is given by $H^{\rm so}= Z\alpha^2 \sum_i{1\over r_i^3} {\bf l_i.s_i}$. The 
 $R$-matrix Breit-Pauli (BPRM) approximation also includes partial two-body 
 interaction terms, such as the ones without momentum operators \cite{aas}. 
 In this approximation, the set of ${SL\pi}$ is re-coupled for the $J\pi$ 
 levels of the (e$^-$ + ion) system which is then followed by diagonalization 
 of the Hamiltonian. 

 The solution of the BPRM approximation is a continuum wave function, $\Psi_F$, 
 for an electron with positive energies (E $>$ 0), or a bound state, $\Psi_B$, 
 with a negative total energy (E $\leq$ 0). The complex resonant structures 
 seen in the photoionization (PI) spectra are produced from channel couplings 
 between continuum channels that are open ($k_i^2~>$ 0), and ones that are 
 closed ($k_i^2~<$ 0), at electron energies $k_i^2$ corresponding to autoionizing 
 states of the Rydberg series, $S_iL_i~J_i\pi_i\nu \ell$. $\nu$ is the effective 
 quantum number of the series converging to excited core thresholds.

 The PI  cross section ($\sigma_{PI}$) is given by

 \begin{equation}
  \sigma_{PI} = {4\pi^2 \over 3c}{1\over g_i}\omega{\bf S} \ ,
 \end{equation}

\noindent 
 where $g_i$ is the statistical weight of the bound state, $\omega$
 is the incident photon energy and ${\bf S}$ is the generalized line strength

\begin{equation}
 {\bf S}= |<\Psi_f||{\bf D}_L||\Psi_i>|^2 =
 \left|\left\langle{\mit\psi}_f \vert\sum_{j=1}^{N+1} r_j\vert
 {\mit\psi}_i\right\rangle\right|^2 \label{eq:SLe},
\end{equation}

\noindent
$\mit\Psi_i$ and $\mit\Psi_f$ are the initial and final state wave
functions, and ${\bf D}_L$ is the dipole operator in length form. 

\section{Computation}

 The computations for the PI  cross-sections were carried out using the 
 {\footnotesize BPRM} package of codes \cite{betal95,np94,znp99}. The computation 
 process consists of a number of stages. The first stage, {\footnotesize STG1}, 
 is initiated with the wave function of the core ion as the initial input. 
 The core ion wave function was obtained from atomic structure calculations 
 using code {\footnotesize SUPERSTRUCTURE} ({\footnotesize SS}) \cite{ejn74,necp03} 
 which uses the Thomas-Fermi-Dirac-Amaldi potential and includes relativistic 
 contributions from the Breit-Pauli approximation. Table \ref{coreNeIV} 
 presents the 58 levels (the ground and 57 excited fine-structure levels) of 
 \ch{Ne^{3+}} included in the wave function expansion of \ch{Ne^{2+}} obtained from 
 the optimization of 17 configurations of the core ion. The table compares 
 the calculated energies from {\footnotesize SS} with the observed values 
 available in the compilation of the NIST \cite{NIST_ASD}, and shows a 
 good agreement between the two. 

\thispagestyle{empty}

\begin{table}
\begin{center}
\caption{\label{coreNeIV} Levels and energies ($E_t$) of core ion \ch{Ne^{3+}} 
in the wave function expansion of \ch{Ne^{2+}}.}
{\scriptsize
\begin{tabular}{rlrll}
\hline
\noalign{\smallskip}
& Level & $J_t$ & $E_t$(Ry) & $E_t$(Ry) \\
&       &       &   NIST    &    SS    \\
 \noalign{\smallskip}
 \hline
 \noalign{\smallskip}
1 & $1s^22s^22p^3(^4S^{\circ})$   & 3/2 & 0.0     & 0. \\
2 & $1s^22s^22p^3(^2D^{\circ})$   & 5/2 & 0.375758&0.4143    \\
3 & $1s^22s^22p^3(^2D^{\circ})$   & 3/2 & 0.37616 &0.4139    \\
4 & $1s^22s^22p^3(^2P^{\circ})$   & 3/2 & 0.5690   &0.5796   \\
5 & $1s^22s^22p^3(^2P^{\circ})$   & 1/2 & 0.5689   &0.5793   \\
6 & $1s^22s2p^4(^4P  )$   & 5/2 & 1.6755  &1.6906     \\
7 & $1s^22s2p^4(^4P  )$   & 3/2 & 1.6811  &1.6961     \\
8 & $1s^22s2p^4(^4P  )$   & 1/2 & 1.6840   &1.6993     \\
9 & $1s^22s2p^4(^2D  )$   & 5/2 & 2.3153   &2.4326     \\
10& $1s^22s2p^4(^2D  )$   & 3/2 & 2.3155   &2.4325      \\
11& $1s^22s2p^4(^2S  )$   & 1/2 & 2.7304   &2.8349     \\
12& $1s^22s2p^4(^2P  )$   & 3/2 & 2.9163   &3.1095     \\
13& $1s^22s2p^4(^2P  )$   & 1/2 & 2.9228   &3.1162     \\
14& $1s^22s^22p^23s(^4P  )$   & 1/2 & 4.3622   &4.3915     \\
15& $1s^22s^22p^23s(^4P  )$   & 3/2 & 4.3657   &4.3949     \\
16& $1s^22s^22p^23s(^4P  )$   & 5/2 & 4.3710   &4.4006     \\
17& $1s^22p^5(^2P^{\circ})$   & 3/2 & 4.4188   &4.6890     \\
18& $1s^22p^5(^2P^{\circ})$   & 1/2 & 4.4275   &4.6986     \\
19& $1s^22s^22p^23s(^2P  )$   & 1/2 & 4.4516   &4.4976     \\
20& $1s^22s^22p^23s(^2P  )$   & 3/2 & 4.4579   &4.5043     \\
21& $1s^22s^22p^23p(^2S^{\circ})$   & 1/2 &          & 4.7018    \\
22& $1s^22s^22p^23s(^2D  )$   & 5/2 & 4.6628   &4.7094     \\
23& $1s^22s^22p^23s(^2D  )$   & 3/2 & 4.6630   &4.7094     \\
24& $1s^22s^22p^23p(^4D^{\circ})$   & 1/2 & 4.7478   & 4.7572    \\
25& $1s^22s^22p^23p(^4D^{\circ})$   & 3/2 & 4.7498   & 4.7592    \\
26& $1s^22s^22p^23p(^4D^{\circ})$   & 5/2 & 4.7530   & 4.7626    \\
27& $1s^22s^22p^23p(^4D^{\circ})$   & 7/2 & 4.7573   & 4.7674    \\
28& $1s^22s^22p^23p(^4P^{\circ})$   & 1/2 & 4.7795   & 4.7965    \\
29& $1s^22s^22p^23p(^4P^{\circ})$   & 3/2 & 4.7812   & 4.7984    \\
30& $1s^22s^22p^23p(^4P^{\circ})$   & 5/2 & 4.7844   & 4.8018    \\
31& $1s^22s^22p^23p(^2D^{\circ})$   & 3/2 & 4.8367   & 4.8643    \\
32& $1s^22s^22p^23p(^2D^{\circ})$   & 5/2 & 4.8569   & 4.8710    \\
33& $1s^22s^22p^23p(^4S^{\circ})$   & 3/2 & 4.8569   & 4.8730    \\
34& $1s^22s2p^33s(^6S^{\circ})$   & 5/2 & 4.9072   & 5.0673    \\
35& $1s^22s^22p^23p(^2P^{\circ})$   & 1/2 &          & 4.9367    \\
36& $1s^22s^22p^23p(^2P^{\circ})$   & 3/2 &          & 4.9372    \\
37& $1s^22s^22p^23s(^2S  )$   & 1/2 & 5.0301   & 5.1169    \\
38& $1s^22s^22p^23p(^2F^{\circ})$   & 5/2 & 5.0602   & 5.0983    \\
39& $1s^22s^22p^23p(^2F^{\circ})$   & 7/2 & 5.0614   & 5.0997    \\
40& $1s^22s^22p^23p(^2D^{\circ})$   & 5/2 & 5.1133   & 5.1863    \\
41& $1s^22s^22p^23p(^2D^{\circ})$   & 3/2 & 5.1143   & 5.1868    \\
42& $1s^22s^22p^23p(^2P^{\circ})$   & 1/2 &          & 5.2463    \\
43& $1s^22s^22p^23p(^2P^{\circ})$   & 3/2 &          & 5.2495    \\
44& $1s^22s^22p^23d(^2P  )$   & 3/2 & 5.2511   & 5.3040    \\
45& $1s^22s^22p^23d(^4F  )$   & 3/2 &          & 5.2679    \\
46& $1s^22s^22p^23d(^4F  )$   & 5/2 &          & 5.2697    \\
47& $1s^22s^22p^23d(^4F  )$   & 7/2 &          & 5.2725    \\
48& $1s^22s^22p^23d(^4F  )$   & 9/2 &          & 5.2761    \\
49& $1s^22s^22p^23d(^2P  )$   & 1/2 & 5.2547   & 5.3116    \\
50& $1s^22s^22p^23d(^2D  )$   & 3/2 & 5.2599   & 5.4149    \\
51& $1s^22s^22p^23d(^2D  )$   & 5/2 & 5.2599   & 5.4168    \\
52& $1s^22s^22p^23d(^4P  )$   & 5/2 & 5.2790   & 5.3275    \\
53& $1s^22s^22p^23d(^4P  )$   & 3/2 & 5.2819   & 5.3300    \\
54& $1s^22s^22p^23d(^4P  )$   & 1/2 & 5.2830   & 5.3313    \\
55& $1s^22s^22p^23d(^4D  )$   & 1/2 &          & 5.3059    \\
56& $1s^22s^22p^23d(^4D  )$   & 5/2 &          & 5.3090    \\
57& $1s^22s^22p^23d(^4D  )$   & 3/2 &          & 5.3092    \\
58& $1s^22s^22p^23d(^4D  )$   & 7/2 &          & 5.3109    \\
            \noalign{\smallskip}
\hline
\hline
\end{tabular}
}
\end{center}
\end{table}

 The wave function included the set of 0 $\leq\ell\leq$ 11 partial waves 
 for the interacting electron and 14 continuum functions for the $R$-matrix 
 basis sets. The $R$-matrix boundary was chosen to be large enough (10 $a_o$) 
 to accommodate the bound orbitals. The second term of the wave function, 
 which represents the bound state correlation functions, included 95 
 ($N$+1)-particle configurations with orbital occupancies from a minimum
 to a maximum number (given within the parentheses) of the orbitals 1s(2-2),
 2s(0-2), 2p(0-6), 3s(0-2), 3p(0-2), 3d(0-2), 4s(0-1), 4p(0-1). 

 PI  cross sections are obtained considering radiation damping 
 for all bound levels using the {\footnotesize BPRM} $R$-matrix codes
 \cite{betal95,np94,znp99}. The narrow PI resonances were delineated 
 at a very fine energy mesh.


\section{Experiment}

 The experimental setup used was located at the Advanced Light Source
 synchrotron at
 the Lawrence Berkeley National Laboratory (LBNL) and  has been extensively described 
 in previous publications. Here we offer a description which is intended to be 
 sufficient to understand the details of this particular study. 
 For more general aspects of the experimental method we refer the interested 
 reader to publications by Covington {\it et al.} \cite{covington2002} and
 by M\"uller and collaborators \cite{muller2015} and also \cite{hinojosa2017}.
 
 The experimental technique consisted in overlapping an ion beam and a photon beam. 
 The ion beam was generated with an electron cyclotron resonance (ECR) ion 
 source. For the photon beam, we used an end-line station of the synchrotron facility. 
 
 The experiment was designed to measure photoionization cross sections.
 In this study, we report on the cross-section of the following reaction,
 
 \begin{equation}\label{rydbergEQ}
 \mbox{Ne}^{2+} + h\nu \longrightarrow \mbox{Ne}^{3+} + e^- \ ,
 \end{equation}
 
 \noindent 
 in which the resulting \ch{Ne^{3+}} (Ne {\footnotesize \ IV}) ions, or {\it photo-ions}, 
 were counted as a function of the photon energy and relevant parameters of 
 the ion beam and the photon beam together with the spatial overlap of the 
 two beams monitored. 
 
 To produce the ion beam of \ch{Ne^{2+}} (Ne {\footnotesize \ III}), a mixture 
 of \ch{Ne} and $Ar$ gases was injected into the all-permanent magnets ECR 
 ion source. A 60$^{\circ}$-sector analyzing magnet was subsequently used 
 to select the Ne$^{2+}$ cations which had a kinetic energy of 12 keV. With 
 the help of an electrostatic Einzel lens, together with two sets of steering 
 plates and a set of slits, the ion beam was focused and collimated. Subsequently, 
 an electrostatic 90$^{\circ}$ spherical-sector analyzer was used to merge 
 the ion beam with the counter-propagating photon beam.
 
 The key section of this experiment was a voltage-biased metal-mesh cylinder 
 where the \ch{Ne^{3+}}-photoions produced inside coupled with a different
 electrostatic potential than the potential experienced by ions produced 
 outside. This metal-mesh cylinder section is defined as the interaction 
 region (IR). The photoions generated inside this IR were separated from 
 those produced outside the IR via a 45$^{\circ}$ dipole analyzing magnet 
 located downstream (as measured by the direction of ion beam travel) from 
 the IR.

 To produce the photon beam, the end line of the synchrotron was equipped 
 with an undulator located inside the ring. The period of the undulator 
 was 10 cm. The synchrotron was operated at 1.9 GeV under a constant electron 
 current of 500 mA. The photon beam was directed onto a spherical 
 grating with controls to scan the photon beam energy and to select the
 energy resolution. 
 The resulting photon beam had a spatial width less than 1.5 mm, a 
 maximum divergence of 0.06$^{\circ}$ and an energy resolution of 8.2 meV. 
 This resolution was estimated from the widths of the resulting resonant 
 peaks in the spectra.
 
 To calibrate the photon energy, an ionization gas-cell was implemented on 
 a side branch of the photon beam line. In this cell, $He$ or $Kr$ gases
 were irradiated in the approximate energy range of 18.6 eV to 92.4 eV. 
 For reference values, we used the data by King {\it et al.} \cite{king1977}
 and Domke and collaborators \cite{domke1996}. The Doppler shift associated
 with the counter-propagating beams was also taken into consideration. 
 An energy uncertainty of $\pm$10 meV was estimated as a result of this
 particular energy calibration procedure.
  
 In the experiment, the resulting \ch{Ne^{3+}} photoions were counted as a 
 function of the photon energy. This procedure resulted in an energy spectrum 
 of the relative intensity of the \ch{Ne^{3+}} signal that was later normalized 
 to absolute cross-section measurements.

 To measure the absolute cross-section, exact knowledge of the interaction
 region dimensions and of the  overlap of the two beams was required. To this end,
 two-dimensional beam profiles of the ion beam $I^+(x,y)$ and the photon beam 
 $I^{\gamma}(x,y)$ were measured. Three beam profilers were used to sample 
 the form factor $F(z)$ according to 

 \begin{equation}
  F_i(z) = \frac{\int\int I^+(x,y)I^{\gamma}(x,y)dxdy}{\int\int I^+(x,y) dxdy  \int\int I^{\gamma}(x,y)dxdy} \ ,
 \end{equation}
 
 \noindent
 where $z$ is the reference axis assigned to the propagation direction of the 
 ion beam. $F(z)$ was measured at three positions; at the entrance, in the 
 center, and at the exit of the IR. 
 These three values, $F_i(z)$, were used to derive $F(z)$ by interpolation along 
 the total IR length and derive, by integration over $z$, the spatial overlap 
 of the photon and ion beams along the common IR path \cite{covington2002}. 
 
 The single-photon photoionization cross-section, $\sigma$, for \ch{Ne^{2+}} was computed 
 from

 \begin{equation}\label{csec}
  \sigma = \frac{Rqe^2v_i\epsilon}{I^+I^{\gamma} \int F(z)dz}  \ ,
 \end{equation}

 \noindent
 where $R$ is the photoion count rate, $q = 2$ is the charge state of \ch{Ne^{2+}},
 $e = 1.6\times10^{-19}$ C, $v_i$ is the ion beam velocity in $cm \cdot s^{-1}$, 
 $\epsilon$ is the responsivity of the photo-diode (electrons per photon), 
 $I^+$ is the ion beam current (A), and $I^{\gamma}$ is the photo-diode current 
 (A).
 
 The \ch{Ne^{2+}} ion beam was measured in an extended Faraday cup. The resulting
 \ch{Ne^{3+}} photoions were counted on a single-channel detector located aft of 
 the dipole analyzing magnet whose magnetic field intensity was adjusted so 
 that only photoions generated inside the IR were collected. A background 
 subtraction of the \ch{Ne^{3+}} signal was performed by cutting the photon beam 
 with a chopper wheel. In this manner,\ch{Ne^{3+}} ions produced by collisions 
 with residual gas were eliminated. 

 The most important sources of systematic error in $\sigma$ in Eq. \ref{csec} 
 came from the beams-overlap integral, the beams profile measurements, and the 
 photo-diode responsivity function. Other contributions to the total systematic 
 error are listed in Covington {\it et al.} \cite{covington2002}. All sources 
 of systematic error combine to a total uncertainty of 20\%.

 The energy spectrum of the relative intensity of \ch{Ne^{3+}} was normalized
 to absolute cross-section measurements that were performed at specific energy
 values of the spectrum at which no resonant peaks were found. A normalization 
 function was derived for these energy values by
 taking the ratio of the cross-section to the relative intensity 
 spectrum.     

 The spectra shown in Fig. \ref{results1} and Fig. \ref{results2} were 
 measured at 5 to 10 eV-wide photon energy intervals. Each interval overlapped 
 its neighboring interval by 1.0 eV. All the individual parts of the spectra 
 were later combined by matching peaks between adjacent intervals.
 An estimation of the maximum possible photon energy uncertainty caused by this 
 procedure was $\pm$8 meV. Therefore, the overall energy 
 uncertainty propagated by both the gas-cell energy calibration and the data 
 reduction procedure is $\pm$13 meV.

\section{Results}

\begin{figure*}
\begin{center}
\includegraphics[width=\textwidth]{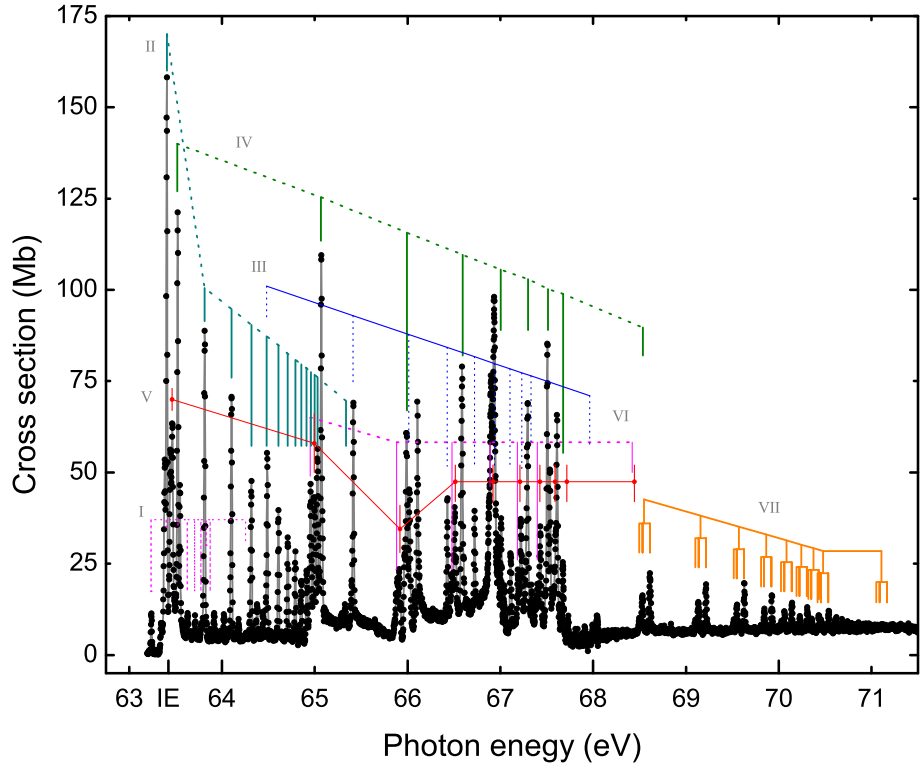}
\caption{\label{results1} (Color online) Single-photon photoionization cross-section 
for Ne$^{2+} + h\nu \rightarrow$ Ne$^{3+} + e^-$. For clarity, the rest of the measured 
spectrum is shown in Fig. \ref{results2}. The identified Rydberg 
series are grouped with inclined lines that intersect with vertical lines. 
The vertical lines indicate the positions of the 
resonances ($E_n$) in the spectrum. The last vertical line of each group
corresponds to the limit $E_{\infty}$. Each group of lines is identified 
with a Roman number in Tables \ref{table-I-III}, \ref{tableIV-VI} and 
\ref{table-VII-VIII}. Label IE indicates the position of the ground state 
ionization energy.
}
\end{center}
\end{figure*}

\begin{figure*}
\begin{center}
\includegraphics[width=\textwidth]{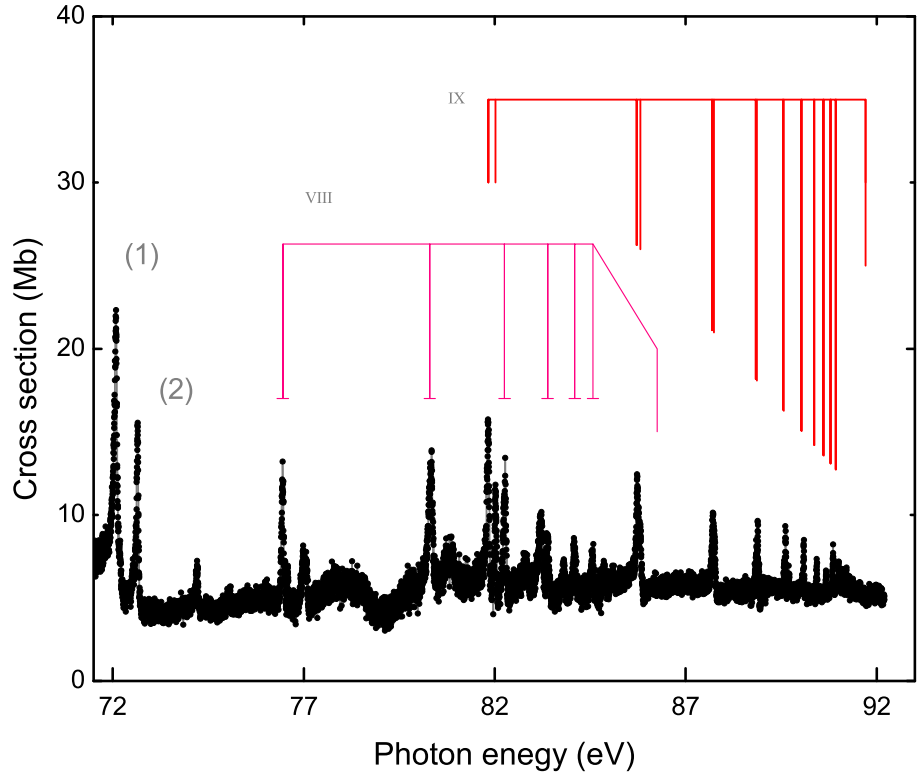}
\caption{\label{results2} (Color online) Single-photon photoionization 
cross-section for Ne$^{2+}$ + $h\nu \rightarrow$ Ne$^{3+} + e^-$. This figure 
is a continuation of Fig. \ref{results1}. Note the vertical scale is 
different. As in Fig. \ref{results1}, identified Rydberg series are grouped 
with horizontal lines that intersect with vertical lines. The vertical lines 
indicate the resonance positions ($E_n$) in the spectrum. The last vertical 
line of each group corresponds to the limit $E_{\infty}$. Groups of lines 
are identified with Roman numbers in Tables \ref{table-VII-VIII}
and \ref{table-IX}. The unidentified peak labelled by (1) has a Gaussian 
center at $72.074$ eV, $I$ = 0.09 $eV\cdot Mb$ and $\omega$ = 63 meV. The 
peak labelled by (2) has a Gaussian center at $72.646$ eV, $I$ = 0.45 
$eV\cdot Mb$ and $\omega$ = 42 meV.
}
\end{center}
\end{figure*}

\begin{figure*}
 \begin{center}
 \includegraphics[width=0.7\textwidth]{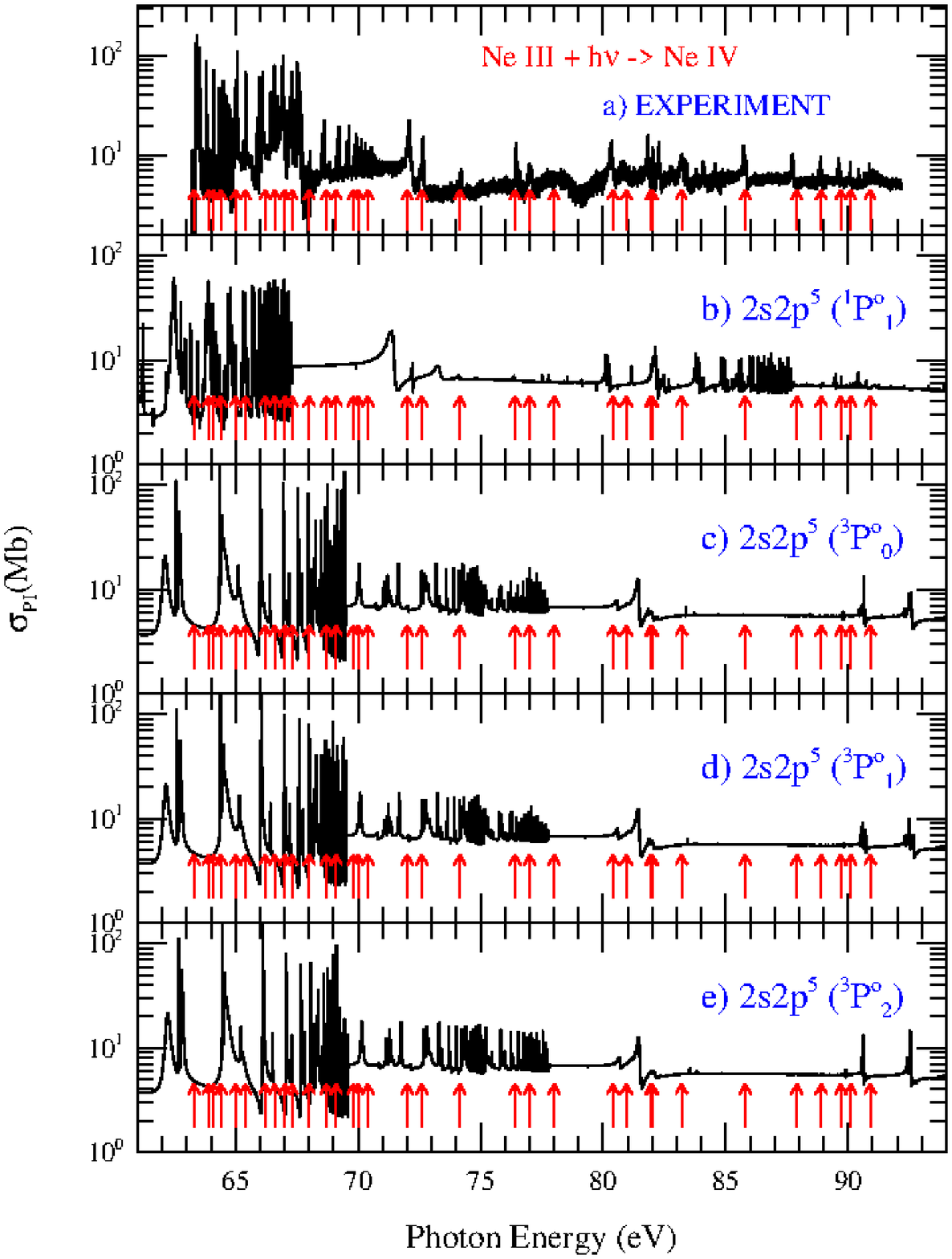}
 \caption{\label{theo3} (Color online) Comparison between the measured and the calculated 
          single-photon photoionization cross sections of \ch{Ne^{2+}}. Note the vertical
          logarithmic scale. The top panel corresponds to the experimental spectrum. 
          The lower panels show the theoretical results for the indicated initial states 
          of \ch{Ne^{2+}}. Arrows indicate the positions of some resonances to assist in 
          their assignments.}
 \end{center}
\end{figure*}

 The measured cross-section is shown in Figs. \ref{results1} and \ref{results2}.
 Results of the calculations for the single-photon PI cross-section of Ne{\footnotesize \ III} 
 are presented in Figs. \ref{theo1}, \ref{theo2} and \ref{theo3}. 
 
 The spectrum consists of several resonant peaks superimposed on a somewhat
 tenuous background. Most of the resonant peaks are attributed to Rydberg series 
 whose energies are given by 
 
 \begin{equation}\label{rydberg}
  E_n = E_{\infty} - \frac{z^2R}{(n - \delta)^2}  \ \ ,
 \end{equation}
  
 \noindent where $E_n$ are the resonant energies, $E_{\infty}$ is the energy
 limit, $z$ is the charge state of ion core (in this case 3), $R$ = 13.606 eV, 
 $n$ is the  principal quantum number and $\delta$ is the quantum defect. 

 The energy limits $E_{\infty}$ in the spectra of Figs. \ref{results1} and 
 \ref{results2} and in Tables \ref{table-I-III}, \ref{tableIV-VI}, 
 \ref{table-VII-VIII} and \ref{table-IX} are computed from the ionization 
 energy of the ground state (indicated by IE in Fig. \ref{results1}), minus 
 the excitation energy of the initial \ch{Ne^{2+}} state, plus the excitation 
 energy of the final \ch{Ne^{3+}} state. Using this derivation for $E_{\infty}$, 
 series were identified. As reference values for the ionization 
 energies, excitation energies and term splittings, we used tabulated NIST 
 values \cite{NIST_ASD}.

 In Figs. \ref{results1} and \ref{results2}, resonant energies $E_n$ are 
 indicated by vertical lines grouped by an inclined or horizontal line according
 to the series to which they are assigned. Roman numbers are used to identify the
 series in the figures and in tables \ref{table-I-III}, \ref{tableIV-VI},
 \ref{table-VII-VIII} and \ref{table-IX}. In each of the series in Figs. 
 \ref{results1} and \ref{results2}, the last vertical line of the group 
 corresponds to the series limit $E_{\infty}$.
 

 Below the ionization energy, one small series was identified as originating 
 from the excited state \ch{Ne^{2+}}, $(2s^22p^4)^1S_0$. This series is labelled 
 by I in Fig. \ref{results1} and in Table \ref{table-I-III}. This series 
 converges to \ch{Ne^{3+}} $(2s^2 2p^3)^2P^{\circ}_{1/2,3/2}$. This doublet $P$ 
 term is split by 1 meV and is unresolved in this experiment.

 Series labelled by II and III originate from the first excited  state 
 of \ch{Ne^{2+}} \ $(2s^22p^4)^1D_2$ and are superimposed over one another. They 
 converge to the doublets \ch{Ne^{3+}} \ $(2s^22p^3)$ \ $^2D^{\circ}_{5/2,3/2}$  
 and  $^2P^{\circ}_{1/2,3/2}$. The energy split of these doublets is 6 meV 
 and 1 meV respectively and could not be resolved by the experiment.
 The only resonance that seems "doubled-peaked" is the first 
 and largest structure of the spectrum and may be attributed to the 
 $^2D_{5/2,3/2}$ splitting.
 Auto-ionizing states of \ch{Ne^{2+}} that may converge to these final states 
 of \ch{Ne^{3+}} are $2s^22p^3(^2D^{\circ}) ns \ ^1D_2^{\circ}$;
 $2s^22p^3(^2D^{\circ}) nd \ ^1F_3^{\circ}, ^1D_2^{\circ}, ^1P_1^{\circ}$;
 $2s^22p^3(^2P^{\circ}) ns \ ^1P_1^{\circ}$ and $2s^22p^3(^2P^{\circ}) nd \ ^1D_2$
 \cite{huffman1967ii}.
 
 In Fig. \ref{results1}, series labelled by IV, V and VI are intermingled. 
 These series are identified as originating from the ground state 
 \ch{Ne^{2+}} $(2s^22p^4)^3P_{2,1,0}$. Their final state seems to be the first 
 excited state of \ch{Ne^{3+}} $(2s^22p^3)^2 D^{\circ}_{5/2,3/2}$. The energy 
 splitting of the final state term is 6 meV and may not be resolved by 
 the experiment. Since these series are 
 overlapped, 
 their assignment as well as the peak energy values are tentative. The series 
 that converge to this final
 state are $2s^22p^3(^2D^{\circ})ns \ ^3D^{\circ}$ and  
 $2s^22p^3(^2D^{\circ})nd \ ^3S^{\circ}, \ ^3D$ \cite{tayal2002}.
 We offer possible assignments to the series labeled by IV, 
 V and VI of Fig. \ref{results1} and Table \ref{tableIV-VI}.

 Below the ionization threshold of \ch{Ne^{2+}} \ $(2s^22p^4)^1D_2$ (66.623 eV), 
 the small background may be caused by the presence of excited \ch{Ne^{2+}} \ 
 $(2s^22p^4)^1S_0$ in the ion beam, and has been identified as series I. The 
 small magnitude of the signal is due to the fact that 
 transitions of the \ch{Ne^{2+}} singlet states $^1D_2$ and $^1S_0$ to 
 \ch{Ne^{3+}} \ $(2s^22p^3)^4S^{\circ}_{3/2}$ are spin-forbidden \cite{huffman1967i}.


 In the energy range of approximately 68 to 71 eV of Fig. \ref{results1}, three 
 Rydberg series labelled by VII have been identified. The first discernible peak 
 of these series corresponds to n = 7. The series can be clearly followed up 
 to n = 15. The largest and clearer series corresponds to the ground state \ch{Ne^{2+}}
 \ $(2s^22p^4)^3P_2$
 converging to the doublet $(2s^22p^3)^2D^{\circ}$ J = 3/2 and 5/2. According to the NIST tables, 
 this doublet is split by 1 meV and therefore cannot be resolved by this experiment. 
 The second series in this energy range corresponds to the next member of the 
 ground state triplet of \ch{Ne^{2+}}$(2s^22p^4)^3P_1$ and, with a lower intensity
 but still somewhat resolved, it appears the last member of the ground state triplet 
 \ch{Ne^{2+}} \ $(2s^22p^4)^3P_0$. Gaussian-fit centers for this series are
 tabulated in Table \ref{table-VII-VIII}. Autoionizing states that may converge
 to these series are 
 $2s^22p^3(^2P^{\circ})ns \ ^3D^{\circ}$ and
 $2s^22p^3(^2P^{\circ})nd \ ^3D^{\circ}, ^3P^{\circ}$ \ .

 Peaks around and about 73 eV of Fig. \ref{results2} have not been associated 
 with a Rydberg
 series. According to our calculations, these peaks are resolved when the initial 
 state of \ch{Ne^{2+}} corresponds to $(2s^22p^4)^1D_2$. Gaussian-fit centers for
 these resonances are given in the Fig. \ref{results2} caption.

  
 In Fig. \ref{results2}, centered approximately at 81 eV, the series labeled by
 VIII originating from the excited state \ch{Ne^{2+}} $(2s^22p^5)^3P_2^{\circ}$, converges 
 to the first $s$ excited state of  \ch{Ne^{3+}} $(2s2p^4)^4P_{5/2,3/2,1/2}$.  
 They all appear as wide peaks in the experiment even when all final and initial 
 states are fully separated.
 In Table \ref{table-IX}, Gaussian-fit centers for this series are presented.

 In Fig. \ref{results2}, resonant structures centered around 87.5 are associated 
 with the final state \ch{Ne^{3+}} $(2s2p^4)^2D_{5/2}$ at 91.721 eV and the initial state
 \ch{Ne^{2+}} \ $(2s^22p^4)^1D_2$.  This series is labelled as series IX. It seems 
 there exists two intermediate states with $\delta$ = 0.40 and 0.48. 
 It was possible to distinguish the peaks at lower energies, but at higher energies
 both series merge. See Table \ref{table-IX}.
   

 The Rydberg series of resonances belonging to core ion excitations are
 relatively well separated, at about 0.37 Ry for $^2D^{\circ}_{5/2,3/2}$, 
 0.57 Ry for $^2P^{\circ}_{3/2,1/2}$ of the ground state configurations, and
 about 1.68 Ry for $^4P_{5/2,3/2,1/2}$ and, 2.3 Ry for $^2D_{5/2,3/2}$ 
 levels (see Table \ref{coreNeIV}). Hence, mixing of resonances belonging 
 to different states is minimal and represents represents one main reason 
 for their accurate identification.
 
 The matching of resonances between the measurement and the theory is found
 to be excellent for Ne {\footnotesize III} (Figs. \ref{theo1}, \ref{theo2}, 
 \ref{theo3}). Some differences are expected,
 particularly in the peaks of the resonances. The peak of the measured
 resonance may depend on: (i) the interference of the autoionizing series of
 resonances, (ii) the relative populations of the ground and excited levels, 
 and (iii) changes in the physical conditions, such as the density and temperature 
 of the experimental ion beam. While condition (iii) has been neglected,
 the precise values of population of the levels remain undetermined.
 However, the agreement between the measurement and theory  
 (Figs. \ref{theo1}, \ref{theo2}, \ref{theo3})
 determines the existence of levels of $2s^22p^4(^3P_{0,1,2},^1D_2,^1S_0)$,
 and $2s2p^5(^1P^{\circ}_1,^3P^{\circ}_{0,1,2})$ in the ion beam and most 
 probably to the statistical distribution.

\subsection{Comparison with experiment}

\begin{figure*}
 \begin{center}
 \includegraphics[width=0.7\textwidth]{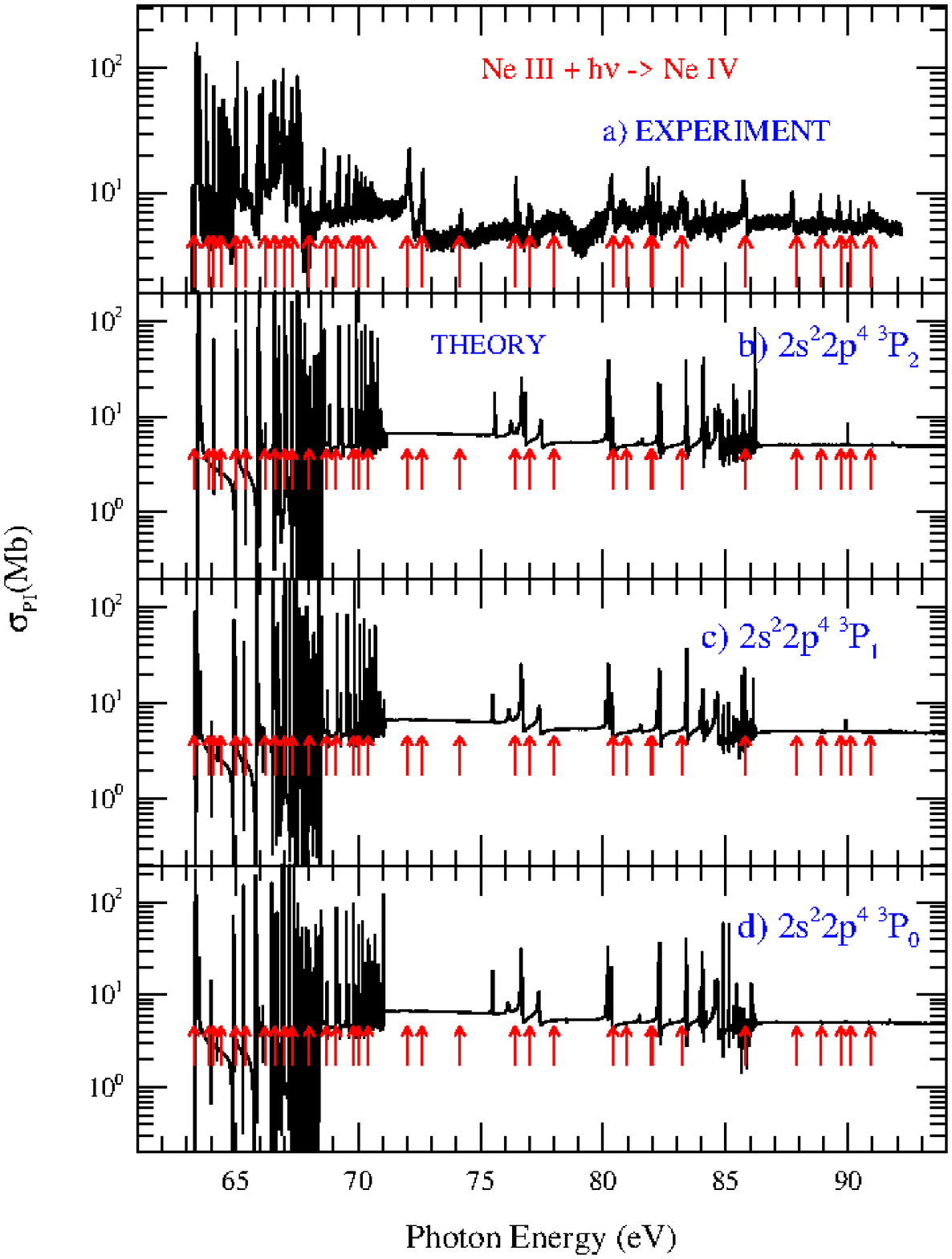}
 \caption{\label{theo1} (Color online) Comparison between the measured and 
          the calculated single-photon photoionization cross-sections of 
          \ch{Ne^{2+}}. Note the vertical logarithmic scale. The top panel 
          corresponds to the experimental spectrum. The lower panels show the 
          theoretical results for the indicated initial states of \ch{Ne^{2+}}. 
          Arrows indicate the positions of some resonances to assist in 
          their assignments.}
 \end{center}
\end{figure*}

\begin{figure*}
 \begin{center}
 \includegraphics[width=0.7\textwidth]{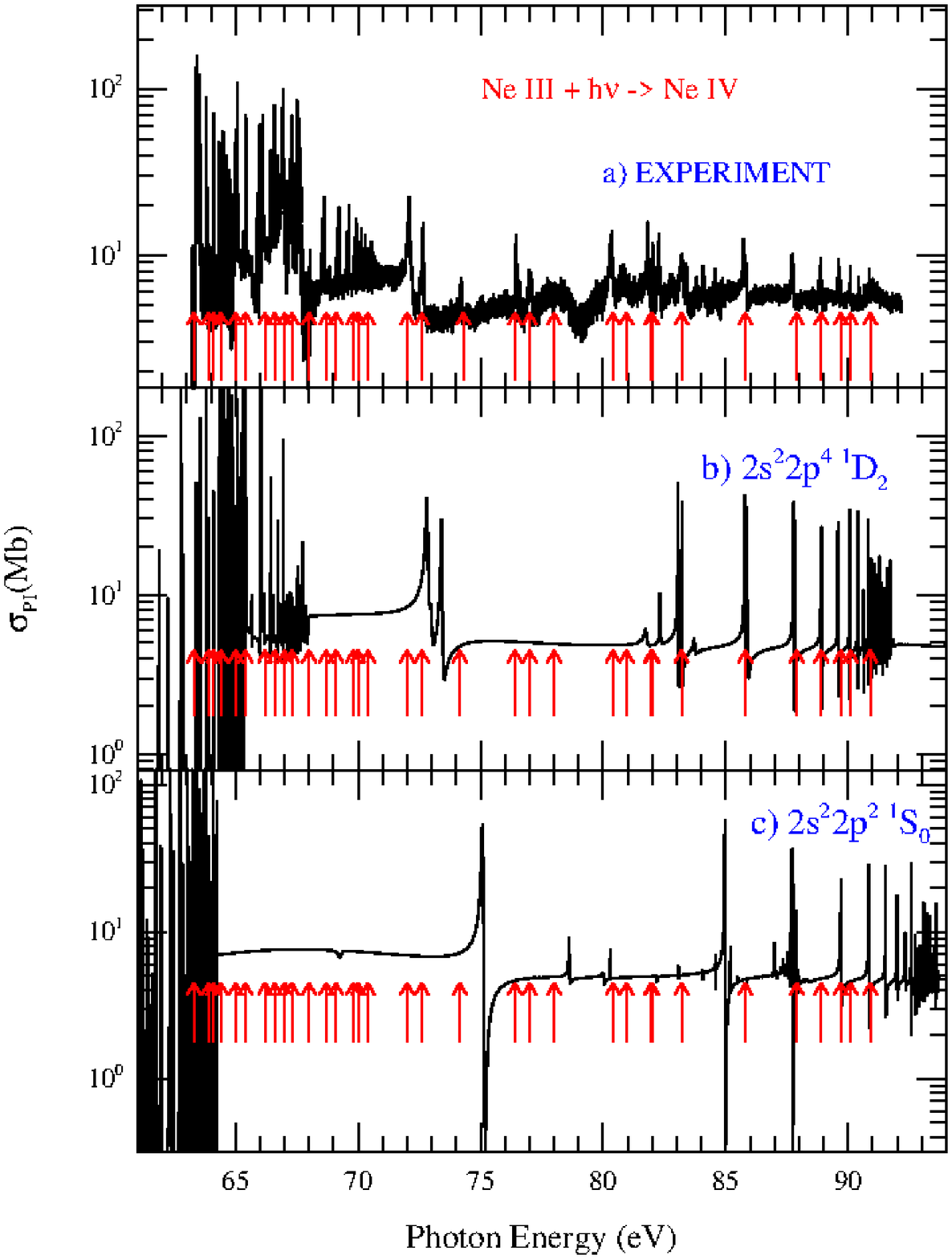}
 \caption{\label{theo2} (Color online) Comparison between the measured and the calculated 
          single-photon photoionization cross-sections of \ch{Ne^{2+}}. Note the vertical
          logarithmic scale. The top panel corresponds to the experimental spectrum. 
          The lower panels show the theoretical results for the indicated initial states 
          of \ch{Ne^{2+}}. Arrows indicate the positions of some resonances to assist in 
          their assignments.}
 \end{center}
\end{figure*}

 Detailed comparisons between the measured and predicted spectrum of $\sigma_{PI}$ 
 is presented in Figs. \ref{theo1}, \ref{theo2} and \ref{theo3}. A remarkable 
 aspect of this experiment, apart from its high resolution, is the close match 
 of the measured spectrum against theory. Resonances appear clearly and allow 
 direct assignments to the theoretically predicted resonances. This makes 
 the manipulation of data through convolution redundant. 

 The measured spectrum has features, integrated from eight low-lying levels, 
 five from the ground configuration $2s^22p^4(^3P_{2,1,0},^1D_2,^1S_0)$ 
 (Fig. \ref{theo1}) and four from the next excited configuration
  $2s2p^5(^3P^{\circ}_{0,1,2},^1P^{\circ}_1)$ (Fig. \ref{theo2}). A number of these 
  levels yield similar features at about the same photon energies. To illustrate 
  the combined features appearing in the observed PI spectrum, the spectrum is 
  divided into three plots: Figs. \ref{theo1}, \ref{theo2} and \ref{theo3}. 
  Most of the prominent resonant structures in the observed spectrum have been 
  marked by arrows in all panels presenting cross-sections of the nine levels 
  in order to properly identify the relevant structures.  

 Fig. \ref{theo1}a compares the observed spectrum with calculated spectra 
 originating from the three levels $^3P_2$ and $^3P_{1,0}$ of the ground state 
 $^3P$. Most of the structures below 68 eV in the observed spectrum are due 
 to Rydberg series of resonances belonging to $2s^22p^3(^2D^2_{5/2,3/2}) n (s,d)$ 
 converging to the residual ion state $^2D^{\circ}$ (lying about 5 eV above 
 the ionization threshold). Comparison with theory shows that most of the 
 observed features come from these three levels. The experimental parameters 
 were set largely in an effort to observe these features. In the lower-energy 
 region, the observed features up to a photon energy of about 71 eV match with 
 those of the ground state beyond which (up to 75 eV) some features are missing. 
 Features in the middle-energy region of about 76 to 86 eV are identified in 
 the predicted cross-sections of the ground state. However, the higher-energy 
 region features in the energy range of 87 to 91 eV (except the one at 90 eV), 
 are also missing. 

 Fig. \ref{theo2} shows the comparison of the observed spectrum with the calculated 
 spectra originating from the other two terms $^1D_2$ and $^1S_0$ of the ground 
 configuration. While resonances similar to those of the state $^3P$ appear for 
 the $^1D_2$ level (and partially for the $^1S_0$ level), and thus contribute 
 to the combined spectrum, missing features in the energy region of 71 to 75 eV 
 of $\sigma_{PI}$ can now be seen in the levels $^1D_2$ and $^1S_0$. Those in 
 the energy region of 71 to 73 eV originate from the $\sigma_{PI}$ of $^1D_2$ 
 and the one at 74.3 eV originates from the $\sigma_{PI}$ of $^1S_0$. They are, 
 however, slightly shifted to the left by the configuration interaction in the 
 theoretical calculations. The missing structures in the high-energy region 
 of 87 to 91 eV  also appear in the $\sigma_{PI}$ for $^1D_2$ and $^1S_0$.  

 Fig. \ref{theo3} compares the observed spectrum with the calculated spectra 
 originating from the four excited levels, $2s2p^5(^1P^{\circ}_1, ^3P^{\circ}_{0,1,2})$, 
 all of which are higher-lying levels compared to those of the ground configuration.  
 Nonetheless they still contribute to the observed spectrum, particularly around 
 the energy interval of 64 to 67 eV. The term $^1P^{\circ}_1$ contributes 
 significantly to the background in the region 68 to 72.5 eV of the observed spectrum.

\section{Conclusions}

 In this study, the single-photon photoionization cross-section of Ne {\footnotesize III}
 leading  to Ne {\footnotesize IV} has been measured in the energy range from 
 below the ground state ionization energy up to 92 eV.  The experimental results 
 have been compared with predictions calculated via the relativistic $R$-matrix 
 Breit-Pauli (BPRM) approximation  The photon energy resolution of the experiment 
 was 8.2 meV. For the theoretical calculations, a close-coupling wave function 
 expansion including 58 fine-structure levels of the residual ion Ne {\footnotesize IV} 
 was included. 

 The spectrum exhibits several peaks, most of which are due to Rydberg series 
 resonances. Identifications and energies for most of these resonances 
 are given. It was possible to identify features originating from lower-lying states of 
 Ne {\footnotesize III} (of the electronic configuration $(2s^22p^4)$, 
 terms $^3P_{210}$, $^1D_{2}$, and $^1S_{0}$), and also from the $(2s2p^5)$ term 
 $^3P^{\circ}_{210}$. In the theoretical calculations, the next excited state of 
 Ne {\footnotesize III}, $(2s2p^5)$ term 
 $^1P^{\circ}_{1}$, was also included, and some peaks associated 
 with this state were clearly identified. As for the core final states 
 identified in the experiment, it was possible to realize 
 all of the first five lower-lying terms of \ch{Ne^{3+}}.
 The present study also found negligible contributions from relativistic effects at 
 the high-energy range of the study.
 
 We hope the present study serves as a benchmark for future photoionization
 applications and theoretical studies of neon ions and 
 oxygen isoelectronic ions.

\section{ Acknowledgements}
 The Advanced Light Source is supported by the Director, Office of Science, 
 Office of Basic Energy Sciences, of the US Department of Energy under 
 Contract No. DE-AC02-05CH11231. A. M. C. acknowledges support through 
 Cooperative Agreement DOE-NA0002075. S. N. N. acknowledges partial support 
 from NSF AST-1312441 and DOE DE-SC0012331 and that the computational work 
 was carried out at the Ohio Supercomputer Center in Columbus Ohio.
 G. H. acknowledges UNAM PAPIIT IN-109-317.

\vspace{-2cm}

\thispagestyle{empty}

 \begin{table} 
 \begin{center}
{\footnotesize
 \begin{tabular}{ c | c | c | c }
\hline 
  $n$  &  $E_n$ &  $I$        &$\omega$  \\
       &  (eV)  &($eV\cdot Mb$)&($meV$)  \\
\hline
\multicolumn{4}{c}{ {\bf I} }   \\ 
\multicolumn{4}{c}{$Ne^{2+}(2s^22p^4)^1S_0 \rightarrow Ne^{3+} (2s^22p^3)^2P_{1/2,3/2}$ } \\
\multicolumn{4}{c}{$\delta=0.02$}\\
\hline
   11   &63.238 &  0.14 &   11.0    \\
   12   &  -    &   -   &    -      \\
   13   &  -    &   -   &    -      \\
   14   &63.629 &  0.08 &   12.0    \\
   15   &63.711 &  0.04 &    7.0    \\
   16   &(63.778)&(0.02)&    -      \\
   17   &63.877 &  0.02 &    8.0    \\
$\vdots$&  -    &   -   &   -       \\
$\infty$&64.253 &   -   &   -       \\
\hline
\multicolumn{4}{c}{ {\bf II} }      \\ 
\multicolumn{4}{c}{ $Ne^{2+}(2s^22p^4)^1D_2 \rightarrow Ne^{3+}(2s^22p^3)^2D_{5/2}$  }  \\
\multicolumn{4}{c}{ $\delta=0.02$ }  \\
\hline
   8   & 63.406 & 1.52  &   9.6      \\
   9   & 63.812 & 1.01  &  11.0      \\
  10   & 64.101 & 1.06  &  13.0      \\
  11   & 64.316 & 0.61  &  12.5      \\
  12   & 64.487 & 0.63  &  11.0      \\
  13   & 64.608 & 0.49  &  11.6      \\
  14   & 64.708 & 0.42  &  12.0      \\
  15   & 64.789 & 0.35  &  12.0      \\
  16   & 64.854 & 0.32  &  15.0      \\
  17   &(64.913)&(0.30) & (13.0)     \\
18$\dagger$&(64.956)&(0.74)& (17.0)  \\
19$\ddagger$&(64.995)&(3.50)& (35.0) \\
  20   &(65.029)&(0.66) & (13.0)     \\
$\vdots$&  -    &   -   &   -        \\
$\infty$&65.332 &   -   &   -        \\
\hline
\multicolumn{4}{c}{ {\bf III} }  \\
\multicolumn{4}{c}{ $Ne^{2+}(2s^22p^4)^1D_2 \rightarrow Ne^{3+}(2s^22p^3)^2P_{1/2,3/2}$  }  \\
\multicolumn{4}{c}{ $\delta=0.07$ }  \\
\hline
   6$^*$&(64.487)&(0.73)&  (12.0)  \\
   7    & 65.414& 0.88  &   12.0   \\
   8    &(66.019)&(0.47)&  (13.0)  \\
   9    & 66.428&  0.42 &   12.0   \\
  10    & 66.721&  0.33 &   11.0   \\
  11    & 67.111&  0.45 &   17.0   \\
12$\dagger\dagger$&(67.221)&(0.35)&(15.0) \\
  13    &(67.335)&(0.09)&   (9.0) \\
$\vdots$&  -    &   -   &   -     \\
$\infty$& 67.962&   -   &   -     \\
\hline
 \end{tabular}   }
 \caption{\label{table-I-III} Gaussian-fit centers ($E_n$) for series labelled 
 by I, II and III in Fig. \ref{results1}. $n$ is the principal quantum number. 
 Oscillator strengths $I$ and widths $\omega$ (approximately 0.849 meV  FWHM) 
 are also given. Values tabulated within brackets are uncertain. Some peaks 
 superimpose between series: $\dagger$ with the $n=6$ peak of series VI; $\ddagger$ 
 with the $n=6$ peak of series V; $^*$with the $n=12$ peaks of series II; $\dagger\dagger$ 
 with the $n=10$ peaks of series V.}
\end{center} 
\end{table} 
\thispagestyle{empty}


\thispagestyle{empty}

 \begin{table} 
 \begin{center}
 {\footnotesize
 \begin{tabular}{c | c | c | c }
\hline 
    $n$    &  $E_n$ &  $I$  & $\omega$       \\
         &  (eV)  &($eV\cdot Mb$)&($meV$) \\
\hline
\multicolumn{4}{c}{ {\bf IV } } \\
\multicolumn{4}{c}{ $Ne^{2+}(2s^22p^4)^3P_2 \rightarrow Ne^{3+}(2s^22p^3)^2D^{\circ}_{5/2} $} \\
\multicolumn{4}{c}{ $\delta=0.05$  } \\
\hline
   5     & 63.522 & 1.38  &  10.0         \\
   6     & 65.069 & 1.12  &  11.0         \\
   7     & 65.989 & 0.54  &  11.0         \\
   8     & 66.586 & 1.16  &  15.7         \\
   9     & 67.006 & 0.30  &  11.3         \\
  10     & 67.292 & 0.87  &  14.4         \\
  11     & 67.506 & 1.05  &  12.0         \\
  12     &(67.673)& 0.31  &  13.6         \\
$\vdots$ &   -    &  -    &   -           \\
$\infty$ & 68.536 &  -    &   -           \\
\hline
\multicolumn{4}{c}{ {\bf V } }\\
\multicolumn{4}{c}{ $Ne^{2+}(2s^22p^4)^3P_1 \rightarrow Ne^{3+}(2s^22p^3)^2D^{\circ}_{5/2,3/2}$} \\
\multicolumn{4}{c}{ $\delta = 0.04$ }      \\
\hline                                     
    5    &(63.463)&(0.20) &  (9.0)         \\
6$\ddagger$    &(64.995)&(3.50) & (35.0)   \\
    7    &(65.908)&(0.11) &  (9.0)         \\
    8    &(66.508)&(0.18) &  (9.0)         \\
    9    &    -   &   -   &    -           \\
10$\dagger\dagger$&(67.221)& (0.35)&(15.0) \\
   11    & 67.429 &  0.35 &   14.0         \\
12$\sharp$    &(67.600)& (0.42)&  (32.0)   \\
   13    & 67.731 &  0.07 &   9.4          \\
$\vdots$ &   -    &   -   &    -           \\
$\infty$ & 68.536 &   -   &    -           \\
\hline
\multicolumn{4}{c}{ {\bf VI } }  \\
\multicolumn{4}{c}{$Ne^{2+}(2s^22p^4)^3P_0  \rightarrow Ne^{3+}(2s^22p^3)^2D^{\circ}_{5/2,3/2} $} \\
\multicolumn{4}{c}{ $\delta = 0.06$ } \\
\hline
6$\dagger$&(64.956)&(0.74) & (17.0) \\
    7    & 65.890 & 0.17  &  13.0   \\
    8    & 66.475 & 0.18  &  14.0   \\
    9    & 67.188 & 0.06  &   7.0   \\
   10    & 67.398 & 0.06  &   7.0   \\
$\vdots$ &   -    &  -    &   -     \\
$\infty$ & 68.422 &  -    &   -     \\
\hline
\end{tabular}  }
\caption{\label{tableIV-VI} Gaussian-fit centers ($E_n$) for series labelled 
 by IV, V and VI in Fig. \ref{results1}. $n$ is the principal quantum number. 
 Oscillator strengths $I$ and widths $\omega$ (approximately 0.849 meV FWHM) 
 are also given. Values tabulated within brackets are uncertain. Some peaks 
 superimpose between series: $\ddagger$ overlaps with the $n=19$ peak of series 
 II. $\dagger\dagger$ overlaps with the $n=12$ peak of series III. $\sharp$ This 
 center was derived from a double Gaussian function fit, the assignment
 corresponds to the smaller peak (see Fig. \ref{results1}) while a large
 peak at 67.612 eV remains unassigned. $\dagger$ Overlaps with series II, 
 $n=18$ peak. }
\end{center}
\end{table}

  \begin{table} 
  \begin{center}
  \begin{tabular}{ c | c | c | c }
 \hline                                    
                                   &        &              &         \\
          $n$                      & $E_n$  &    $I$       &$\omega$ \\
                                   &  (eV)  &($eV\cdot Mb$)&($meV$)  \\ 
 \hline
 \multicolumn{4}{c}{ {\bf VII } }  \\
 \multicolumn{4}{c}{  $Ne^{2+}(2s^22p^4)^3P_2 \rightarrow Ne^{3+}(2s^22p^3)^2P^{\circ}_{1/2,3/2}$  }  \\
 \multicolumn{4}{c}{  $\delta$ = 0.075  }  \\
 \hline
           7                       & 68.611 &    0.32      &   17    \\
           8                       & 69.215	&    0.22      &   14    \\
           9                       & 69.628 &    0.18      &   13    \\                   
           10                      & 69.924 &    0.14      &   12    \\                   
           11                      & 70.141 &    0.10      &   11    \\
           12                      &(70.307)&   (0.08)     &  (13)   \\ 
           13                      &(70.433)&   (0.05)     &  (13)   \\
           14                      &(70.538)&   (0.08)     &  (13)   \\	                                     
        $\vdots$                   &   -    &     -        &    -    \\  
        $\infty$                   & 71.165 &     -        &    -    \\                     
 \hline
 \multicolumn{4}{c}{  $Ne^{2+}(2s^22p^4)^3P_1 \rightarrow Ne^{3+}(2s^22p^3)^2P^{\circ}_{1/2,3/2}$  }  \\
 \hline
           7                       & 68.531 &    0.19      &   16     \\
           8                       & 69.134 &    0.17      &   16     \\                 
           9                       & 69.548 &    0.12      &   13     \\
          10                       & 69.844 &    0.08      &   12     \\
          11                       & 70.061 &    0.01      &   11     \\
          12                       & 70.226 &    0.05      &   11     \\
          13                       &(70.355)&   (0.05)     &  (12)    \\
        $\vdots$                   &   -    &     -        &    -     \\                                      
        $\infty$                   & 71.085 &     -        &    -     \\
 \hline
 \multicolumn{4}{c}{  $Ne^{2+}(2s^22p^4)^3P_0 \rightarrow Ne^{3+}(2s^22p^3)^2P^{\circ}_{1/2,3/2}$  }  \\
 \hline
           7                       & 68.497 &     0.04     &   12   \\
           8                       & 69.100 &     0.02     &   11   \\
           9                       & 69.515 &     0.05     &   14   \\
          10                       &(69.810)&    (0.06)    &  (22)  \\
        $\vdots$                   &   -    &       -      &    -   \\                     
        $\infty$                   & 71.051 &       -      &    -   \\
\hline
\end{tabular}
\caption{\label{table-VII-VIII} Gaussian-fit centers ($E_n$) for series 
labelled by VII in Fig. \ref{results1} and VIII in Fig. \ref{results2}. 
$n$ is the principal quantum number. Oscillator strengths $I$ and widths 
$\omega$ (approximately 0.849 meV FWHM) are also given. Values tabulated 
within brackets are uncertain.}
\end{center}
\end{table}



 \begin{table} 
 \begin{center}
 \begin{tabular}{ c | c | c | c }
\hline
   $n\ell$                  & $E_n$  &    $I$       &$\omega$  \\
                            &  (eV)  &($eV\cdot Mb$)&($meV$)   \\
 \hline
 \multicolumn{4}{c}{ {\bf VIII } }  \\
 \multicolumn{4}{c}{  $Ne^{2+}(2s2p^5)^3P^{\circ}_2  \rightarrow  Ne^{3+}(2s2p^4)^4P $  }  \\
 \multicolumn{4}{c}{ $\delta$ = 0.47 }  \\
 \hline
                                   &        &              &        \\
          4                        & 76.446 &    0.50      &   56   \\
          5                        & 80.321 &    0.71      &   92   \\
          6                        & 82.259 &    0.62      &   75   \\
          7                        & 83.384 &    0.24      &   65   \\
          8                        & 84.083 &    0.20      &   64   \\
          9                        & 84.552 &    0.14      &   62   \\
        $\vdots$                   &   -    &       -      &    -   \\
        $\infty$                   & 86.254 &       -      &    -   \\
\hline
\multicolumn{4}{c}{ {\bf IX }}  \\
\multicolumn{4}{c}{ $Ne^{2+}(2s^22p^4)^1D_2  \rightarrow  Ne^{3+}(2s2p^4)^2D_{5/2,3/2}$  }  \\
\multicolumn{4}{c}{  $\bar{\delta}$ = 0.46$^{\gamma}$  }  \\
\hline
        4 ($\delta$ = 0.44)     &  81.823  &   0.56   &   53   \\
        4 ($\delta$ = 0.48)     &  82.020  &   0.33   &   46   \\
        5 ($\delta$ = 0.44)     & (85.735) &  (0.52)  &  (70)  \\                                      
        5 ($\delta$ = 0.48)     & (85.816) &  (0.15)  &  (42)  \\
        6                       & (87.718) &  (0.44)  &  (82)  \\                   
        7                       & (88.878) &  (0.12)  &  (45)  \\
        8                       & (89.608) &  (0.13)  &  (39)  \\ 
        9                       & (90.089) &  (0.06)  &  (18)  \\
       10                       & (90.430) &  (0.05)  &  (21)  \\
       11                       & (90.675) &  (0.03)  &  (22)  \\
       12                       & (90.858) &  (0.06)  &  (23)  \\
       13                       & (91.004) &  (0.03)  &  (18)  \\                                      
     $\vdots$                   &    -     &     -    &    -   \\  
     $\infty$                   &  91.723  &     -    &    -   \\                     
\hline 
 \end{tabular}
 \caption{
 \label{table-IX} Gaussian-fit centers ($E_n$) for series labelled by IX in 
 Fig. \ref{results2}. $n$ is the principal quantum number. Oscillator strengths 
 $I$ and widths $\omega$ (approximately 0.849 meV FWHM) are also given.
 Values tabulated within brackets are uncertain. $^{\gamma}$This value 
 of $\delta$ corresponds to an average of both series.}
 \end{center}
 \end{table}

\pagebreak
 
\bibliography{k-shell,biblioNEIII,P-biblio-GMO}

\end{document}